\newcommand{\be}{\begin{eqnarray}}
\newcommand{\ee}{\end{eqnarray}}
\def\sgn{\mbox{sgn}}

\def\r#1{\right#1}
\def\fr{\frac{1}{2}}
\def\mref#1{(\ref{#1})}

\def\rr{\mbox{\Bbb R}}

\def\bd{\begin{displaymath}}
\def\ed{\end{displaymath}}

\def\ba#1{\begin{array}{#1}}
\def\ea{\end{array}}
\def\nn{\nonumber}
\newfont{\Bbb}{msbm10 scaled 1200}
\documentclass[12pt]{article}
\usepackage{fleqn}
\title{
On the Calogero model with negative harmonic term
}
\author{
C.Gonera\thanks{supported by KBN grant 2 P03B 076 10}\\
M.Majewski$^*$\\
P.Ma\'slanka$^*$\\
                              Theoretical Physics Department II\\
                                     University of Lodz\\
                          ul. Pomorska 149/153, 90 236 Lodz, Poland\\
}
\date{}
\begin{document}
\maketitle
\thispagestyle{empty}
\begin{abstract}
The Calogero model with negative harmonic term is shown to be equivalent
to the
set of negative harmonic oscillators. Two time--independent canonical
transformations relating both models are constructed: one based on the
recent
results concerning quantum Ca\-lo\-ge\-ro model and one obtained from
dynamical
$sL(2,\rr)$\ algebra. The two--particle case is discussed in some
detail.
\end{abstract}
\section{Introduction \label{s1}}
It has been shown recently~\cite{b1} that the quantum Calogero
model~\cite{b2}
can be transformed by a similarity transformation into the set of 
noninteracting harmonic oscillators. It is also known~\cite{b3}\cite{b4}
that
the Calogero--Moser model (i.e. the model with purely inverse square 
interaction) is equivalent to the set of free particles, both on
classical
and quantum levels.

In the present paper we show that the ``inverted" classical Calogero
model, 
i.e. the model obtained from the Calogero model by changing the sign of
the
oscillator term is equivalent to the set of inverted harmonic
oscillators.
The equivalence is proven by constructing, more or less explicitly, the
relevant time--independent canonical transformation.

Two canonical transformations doing the job are presented. First, we
construct
the classical counterpart of the similarity transformation proposed in 
Ref.\cite{b1}. It is easy to check that for the inverted case this 
transformation is unitary which, in classical case, corresponds to real 
canonical transformation. In sec.\ref{s3} an alternative mapping is
presented
which is based on $sL(2,\rr)$\ dynamical algebra of rational Calogero
model. In order to show that they are really different transformation
we consider in some detail the $N=2$\ case. Both mappings are here
constructed explicitly which makes their inequivalence transparent.

\section{Equivalence of the models \label{s2}}

Let us start with ``inverted" Calogero model described by the
hamiltonian
\be
H_C&=&\sum\limits_i\left({p^2_i\over 2}-{\omega^2q^2_i\over 2}\r)+g/2
\sum\limits_{i\not=j}{1\over (q_i-q_j)^2}\label{w1}
\ee

In order to get rid of the last term on the right hand side we perform
three successive canonical transformations.\hfill{}\\
(i)\ \ First we put
\be
q_i&=&q_i'\label{w2}\\
p_i&=&p_i'+\omega q_i'\nn
\ee
This is a canonical transformation and transforms $H_C$\ into
\be
H_C&=&\sum\limits_i{p_i^{\prime 2}\over 2}+g/2\sum\limits_{i\not=j}
{1\over(q_i'-q_j')^2}+\omega\sum\limits_iq_i'p_i'\label{w3}
\ee
(ii)\ \ Next, consider the one--parameter family of canonical
tramsformations
of the form
\be
q_i'(\lambda)&=&e^{-\lambda H_{CM}(q'',p'')}*q_i''\equiv
q_i''+\lambda\{q_i'',
H_{CM}(q'',p'')\}+\ldots\label{w4}\\
p_i'(\lambda)&=&e^{-\lambda H_{CM}(q'',p'')}*p_i''\equiv
p_i''+\lambda\{p_i'',
H_{CM}(q'',p'')\}+\ldots\nn
\ee
where $H_{CM}$\ is Calogero--Moser hamiltonian
\be
H_{CM}(q,p,)&=&\sum\limits_i{p_i^2\over 2}+g/2\sum\limits_{i\not=j}
{1\over (q_i-q_j)^2}\label{w5}
\ee

Actually, eq.\mref{w4} describes the dynamics of Calogero--Moser model, 
$\lambda$\ playing the role of time variable. It is easy to check that
the
chioce $\lambda=-{1\over 2\omega}$\ gives
\be
e^{{1\over
2\omega}H_{CM}(q'',p'')}*\left(H_{CM}(q'',p'')+\omega\sum\limits_i
q_i''p_i''\r)&=&\omega\sum\limits_iq_i''p_i''\label{w6}
\ee

Therefore, as a second canonical transformation we take
\be
q_i'&=&e^{{1\over 2\omega}H_{CM}(q'',p'')}*q_i''\label{w7}\\
p_i'&=&e^{{1\over 2\omega}H_{CM}(q'',p'')}*p_i''\nn
\ee
implying
\be
H_C&=&\omega\sum\limits_iq_i''p_i''\label{w8}
\ee
(iii)\ \ Finally, let
\be
q_i''&=&{1\over\sqrt{2}}(Q_i+{1\over\omega}P_i)\label{w9}\\
p_i''&=&{1\over\sqrt{2}}(P_i-\omega Q_i)\nn
\ee
so that
\be
\omega\sum\limits_iq_i''p_i''&=&\sum\limits_i\left({P^2_i\over 2}-
{\omega^2Q_i^2\over 2}\r)\label{w10}
\ee

The transformation $(q,p)\to(Q,P)$\ given by the composition of
(i)$\div$(iii)
reduces ``inverted" Calogero model to the inverted harmonic oscillator.

\section{An alternative formulation\label{s3}}

It is well known that the rational Calogero model posses dynamical
$sL(2,\rr)$\ 
symmetry. In fact, define 
\be
T_+&\equiv&{1\over\omega}\left(\sum\limits_ip_i^2/2+g/2\sum\limits_{i\not=j}
{1\over(q_i-q_j)^2}\r)\nn\\
T_-&\equiv&\omega\sum\limits_iq_i^2/2\label{w11}\\
T_0&\equiv&\fr\sum\limits_iq_ip_i\nn
\ee
where $\omega$\ is a fixed, nonzero but otherwise arbitrary, frequency.
The 
relevant Poisson bracket read
\be
\{T_0,T_\pm\}&=&\pm T_\pm\label{w12}\\
\{T_+,T_-\}&=&-2T_0.\nn
\ee
Define
\be
T_1&=&\fr(T_++T_-)\nn\\
T_2&=&\fr(T_+-T_-)\label{w13}\\
T_3&=&iT_0;\nn
\ee
the commutation rules~\mref{w12} take the form
\be
\{T_i,T_j\}&=&i\epsilon_{ijk}T_k\label{w14}
\ee

Consider now the canonical transformation
\be
q_i(\lambda)&=&e^{\lambda T_1(q',p')}*q_i'\label{w15}\\
p_i(\lambda)&=&e^{\lambda T_1(q',p')}*p_i';\nn
\ee
then
\be
T_2(q(\lambda),p(\lambda))&=&e^{\lambda
T_1(q',p')}*T_2(q',p')=\label{w16}\\
&=&T_2(q',p')\cos\lambda+T_3(q',p')\sin\lambda\nn
\ee

Therefore, puting
\be
q_i&=&e^{{\pi\over 2}T_1(q'',p'')}*q_i''\label{w17}\\
p_i&=&e^{{\pi\over 2}T_1(q'',p'')}*p_i''\nn
\ee
one obtains
\be
T_2(q,p)&=&T_3(q'',p'')\label{w18}
\ee
Let us define
\be
\tilde{T}_i&=&T_i(g=0)\label{w19}
\ee
Then $T_3=\tilde{T}_3$\ and the transformation
\be
q_i''&=&e^{-{\pi\over 2}\tilde{T}_1(Q,P)}*Q_i\label{w20}\\
p_i''&=&e^{-{\pi\over 2}\tilde{T}_1(Q,P)}*P_i\nn
\ee
converts $\tilde{T}_3=T_3$\ into $\tilde{T}_2$,
\be
\tilde{T}_2(P,Q)&=&\tilde{T}_3(q''.p'')\label{w21}
\ee

Composing the mappings~\mref{w17} and~\mref{w20} one gets finally
\be
H_C&=&\sum\limits_i\left({P_i^2\over 2}-{\omega^2Q_i^2\over
2}\r)\label{w22}
\ee

The canonical transformation~\mref{w20}, written out explicitly, reads
\be
q_i&=&{1\over\sqrt{2}}(Q_i-{1\over\omega}P_i)\label{w23}\\
p_i&=&{1\over\sqrt{2}}(P_i+\omega Q_i)\nn
\ee
and coincides with eq.\mref{w9}.

\section{The $N=2$\ case\label{s4}}

In secs.\ref{s2} and~\ref{s3} we defined two canonical transformations
reducing
the inverted Calogero hamiltonian to dilatation. However, we still do
not know
whether these are different transformations or two forms of the same 
transformation. To show that they are really different it is sufficient
to
consider the $N=2$\ case. Separating the center of mass and relative 
coordinates
\be
X=\fr(q_1+q_2)&,&\pi=p_1+p_2\label{w24}\\
q=q_1-q_2&,&p=\fr(p_1-p_2)\nn
\ee
and ignoring the former we get
\be
H_C&=&p^2+g/q^2-{\omega^2\over 4}q^2.\label{w25}
\ee

The transformations described in secs.\ref{s2} and~\ref{s3} can be given
explicitly. In particular, from eqs.\mref{w2} and~\mref{w7} one gets
\be
q&=&\sgn(q'')\sqrt{g+(q''p''-E/\omega)^2\over E}\nn\\
p&=&\sgn(q''){(q''p''-E/\omega)\over\sqrt{g+(q''p''-E/\omega)^2\over
E}}+
\sgn(q''){\omega\over 2}\sqrt{g+(q''p''-E/\omega)^2\over E}\label{w26}\\
E&=&p^{\prime\prime 2}+g/q^{\prime\prime 2}\nn
\ee
We easily check that
\be
p^2+g/q^2-{\omega^2\over 4}q^2&=&\omega q''p''\label{w27}
\ee
as it should be.

On the other hand eq.\mref{w17} takes the form
\be
q&=&\sgn(q'')\sqrt{{2E\over\omega^2}-{2q''p''\over\omega}}\nn\\
p&=&\sgn(q''){(\omega q^{\prime\prime 2}/2-E/\omega)\over\sqrt{%
{2E\over\omega^2}-{2q''p''\over\omega}}}\label{w28}\\
E&=&p^{\prime\prime 2}+g/q^{\prime\prime 2}+{\omega^2\over
4}q^{\prime\prime 2}
\nn
\ee
Again one easily verifies that the eq.\mref{w27} holds. By comparying
eqs.\mref{w26} and~\mref{w28} we conclude that the transformations
described in secs.\ref{s2} and~\ref{s3} are really different.


\begin{thebibliography}{99}
\bibitem{b1}
N.Gurappa, P.K.Panigrahi, preprint cond--mat/9710035\\
A.Nishino, U.Ujino, M.Wadati, preprint cond--mat/9803284
\bibitem{b2}
F.Calogero, J.Math.Phys. {\bf 12} (1971), 419
\bibitem{b3}
A.P.Polychronakos, Nucl.Phys. {\bf B324} (1989), 597
\bibitem{b4}
T.Brzezi\'nski, C.Gonera, P.Ma\'slanka, preprint hep--th/9810176
\end{thebibliography}
\end{document}